\documentclass[aps,superscriptaddress,nofootinbib,showpacs,twocolumn]{revtex4} 
\usepackage{graphicx,epsfig,wrapfig,amssymb,epstopdf}
\newcommand{\be}{\begin{equation}}
\newcommand{\ee}{\end{equation}}
\newcommand{\ba}{\begin{eqnarray}}
\newcommand{\ea}{\end{eqnarray}}
\newcommand{\la}{\langle}
\newcommand{\ra}{\rangle}
\newcommand{\di}{ {\rm d} }
\newcommand{\SIDIS}{{\mbox{\tiny SIDIS}}}
\newcommand{\DY   }{{\mbox{\tiny DY}}}
\begin{document}
\newcommand*{\PennState}{Penn State University, 104 Davey Lab, 
University Park PA 16802, U.S.A.}\affiliation{\PennState}
\newcommand*{\Dubna}{Joint Institute for Nuclear Research, Dubna, 
141980 Russia}\affiliation{\Dubna}
\newcommand*{\Bochum}{Institut f{\"u}r Theoretische Physik II, 
Ruhr-Universit{\"a}t Bochum, D-44780 Bochum, Germany}\affiliation{\Bochum}
\newcommand*{\UIUC}{University of Illinois, Urbana, IL 61801, USA}\affiliation{\UIUC}
\newcommand*{\BNL}{RIKEN BNL Research Center, 
Upton, New York 11973}\affiliation{\BNL}

\title{Sivers effect in Drell Yan at BNL RHIC}
\author{J.~C.~Collins}\affiliation{\PennState}\affiliation{\Bochum}
\author{A.~V.~Efremov}\affiliation{\Dubna}
\author{K.~Goeke}\affiliation{\Bochum}
\author{M.~Grosse Perdekamp}\affiliation{\UIUC}\affiliation{\BNL}
\author{S.~Menzel}\affiliation{\Bochum}
\author{B.~Meredith}\affiliation{\UIUC}
\author{A.~Metz}\affiliation{\Bochum}
\author{P.~Schweitzer}\affiliation{\Bochum}
\date{April 2006}


\begin{abstract}
\noindent
On the basis of a fit to the Sivers effect in deep-inelastic
scattering, we make predictions for single-spin asymmetries in the
Drell-Yan process at BNL RHIC.
\end{abstract}


\maketitle

\section{Introduction}
\label{Sec-1:introduction}

The Sivers effect \cite{Sivers:1989cc} was originally suggested 
to explain the large single spin asymmetries (SSA) observed in 
$p^\uparrow p\to\pi X$ (and $\bar{p}^\uparrow p\to\pi X$) at FNAL 
\cite{Adams:1991rw} and recently at higher energies in the RHIC 
experiment \cite{Adams:2003fx}.
The effect considers a non-trivial correlation between (the transverse 
component of) the nucleon spin ${\bf S}_{\rm T}$ and intrinsic transverse 
parton momenta ${\bf p}_{\rm T}$ in the nucleon. It is proportional to the 
``(naively or artificially) T-odd'' structure 
$({\bf S}_{\rm T}\times{\bf p}_{\rm T}){\bf P}_N$ and quantified in terms 
of the Sivers function $f_{1T}^\perp(x,{\bf p}_T^2)$ \cite{Boer:1997nt}
whose precise definition in QCD was worked out only recently 
\cite{Brodsky:2002cx,Collins:2002kn,Belitsky:2002sm}.

A particularly interesting feature of the Sivers function (and other
``T-odd'' distributions) concerns the universality property. On the basis 
of time-reversal arguments it was predicted \cite{Collins:2002kn} that 
$f_{1T}^\perp$ in
semi-inclusive deeply inelastic scattering (SIDIS) and in the Drell-Yan process (DY) 
have opposite sign (our definition of the 
Sivers function follows the Trento Conventions \cite{Bacchetta:2004jz}),
\be\label{Eq:01}
     f_{1T}^\perp(x,{\bf p}_T^2)_\SIDIS =
    -f_{1T}^\perp(x,{\bf p}_T^2)_\DY \;.
\ee
The experimental check of Eq.~(\ref{Eq:01}) would provide a thorough test
of our understanding of the Sivers effect within QCD. In particular, the 
experimental verification of (\ref{Eq:01}) is a crucial prerequisite for 
testing the factorization approach to the description of processes containing
$p_T$-dependent correlators \cite{Collins:1981uk,Ji:2004wu,Collins:2004nx}.

Recent data on SSA from SIDIS \cite{Avakian:1999rr,Airapetian:1999tv,Avakian:2003pk},
and in particular those from transversely polarized targets
\cite{HERMES-new,Airapetian:2004tw,Alexakhin:2005iw},
provide first measurements of the Sivers effect in SIDIS.
On the basis of this information it was shown that the Sivers effect 
leads to sizeable SSA in $p^\uparrow\pi^-\to l^+l^-X$, which could be studied 
at COMPASS \cite{COMPASS-proposal}, and in $p^\uparrow\bar{p}\to l^+l^-X$ or 
$p\bar{p}^\uparrow\to l^+l^-X$ in the planned PAX experiment at GSI 
\cite{PAX,PAX-estimates}, making the experimental check of Eq.~(\ref{Eq:01}) 
feasible and promising \cite{Efremov:2004tp}.
Both experiments, which could be performed in the medium or long term,
have the advantage of being dominated by annihilation of valence quarks
(from $p$) and valence antiquarks (from $\bar{p}$ or $\pi^-$), which yields
sizeable counting rates. Moreover, the processes are not very
sensitive to the Sivers 
antiquark distributions, which are not constrained by the present SIDIS data, 
see \cite{Efremov:2004tp,Collins:2005ie,Anselmino:2005nn,Anselmino:2005ea,Vogelsang:2005cs}.

On a shorter term the Sivers effect in DY can be studied at RHIC in 
$p^\uparrow p\to l^+l^-X$. 
Similar and other spin physics prospects at RHIC are discussed in
\cite{Bunce:2000uv,Boer:1999mm,Boer:2003tx,Dressler:1999zv,Boer:2001tx,Anselmino:2002pd,Anselmino:2004nk}. 
Other earlier predictions for SSAs in DY were given in 
\cite{Boros:1993ps,Hammon:1996pw}.

In $pp$-collisions inevitably antiquark distributions are involved, 
and the counting rates are smaller. In this work we shall demonstrate 
that the Sivers effect SSA in DY can nevertheless be measured at RHIC 
with an accuracy sufficient to unambiguously test Eq.~(\ref{Eq:01}).
In particular, by focusing on certain kinematic regions the effect of the 
unknown Sivers antiquark distribution function can be minimized. 
And, by focusing on the opposite kinematic regions one can gain first 
information on the Sivers antiquark distribution itself.
For our estimates we use the Sivers function extracted from HERMES data 
\cite{Airapetian:2004tw} in \cite{Collins:2005ie}, 
see also~\cite{Collins:2005wb,Anselmino:2005an}.

It remains to be noted that the theoretical understanding of SSA in
$p^\uparrow p\to\pi X$, which originally motivated the introduction of the 
Sivers effect, is more involved and less lucid compared to SIDIS or DY, 
as here also other mechanisms such as the Collins effect \cite{Collins:1992kk} 
and/or dynamical twist-3 effects \cite{Efremov:eb,Kanazawa:2000hz} 
could generate SSA. Phenomenological studies indicate, however, 
that in a picture based on $p_T$-dependent correlators the data 
\cite{Adams:1991rw,Adams:2003fx} can be explained in terms of the Sivers 
effect alone \cite{Anselmino:1994tv,Anselmino:1998yz,D'Alesio:2004up}
with the other effects playing a less important role 
\cite{Anselmino:2004ky,Ma:2004tr}.
For recent discussions of hadron-hadron collisions with more complicated 
final states (like, e.g., $p^\uparrow p\to\mbox{jet}_1\mbox{jet}_2 X$) 
we refer to Refs.\ \cite{Bomhof:2004aw}.

\section{\boldmath Sivers effect in SIDIS}
\label{Sec-2:Sivers-in-SIDIS}

The longitudinal SSA in SIDIS observed first 
\cite{Avakian:1999rr,Airapetian:1999tv,Avakian:2003pk} unfortunately 
cannot be unambiguously interpreted in terms of a unique 
(Sivers \cite{Sivers:1989cc}, Collins \cite{Collins:1992kk} or 
twist-3 \cite{Afanasev:2003ze}) effect 
\cite{Efremov:2001cz,DeSanctis:2000fh,Ma:2002ns,Efremov:2003tf,Anselmino:2004ht}, 
see Ref.~\cite{Efremov:2004hz} for a recent review. All that is clear 
at the present stage is that these SSAs are dominated by subleading twist 
effects \cite{Airapetian:2005jc}.
The situation changed, however, with the transverse SSA in SIDIS observed 
at HERMES and COMPASS \cite{HERMES-new,Airapetian:2004tw,Alexakhin:2005iw},
where the Sivers and Collins effect can be distinguished due to the different 
azimuthal angle distribution of the produced hadrons \cite{Boer:1997nt}.
 
In particular, in the SSA due to the Sivers effect the produced hadrons 
exhibit an azimuthal distribution $\propto\sin(\phi-\phi_S)$ in the plane 
transverse to the beam axis.
Here $\phi$ and $\phi_S$ are the azimuthal angles of respectively, 
the produced hadron and of the target polarization vector, 
with respect to the lepton scattering plane.

The Sivers SSA measured at HERMES \cite{Airapetian:2004tw} 
is defined  as sum over SIDIS events $i$ as 
\ba\label{Eq:AUT-Siv-unw-SIDIS-exp}
 && A_{UT}^{\sin(\phi-\phi_S)}      \\
 && = \frac{\sum_i\sin(\phi_i-\phi_{S,i})
    \{N^\uparrow(\phi_i,\phi_{S,i})-N^\downarrow(\phi_i,\phi_{S,i}+\pi)\}}
    {\frac12\sum_i
    \{N^\uparrow(\phi_i,\phi_{S,i})+N^\downarrow(\phi_i,\phi_{S,i}+\pi)\}}
        \;.\nonumber\ea
$N^{\uparrow(\downarrow)}(\phi_i,\phi_{S,i})$ are the event counts for 
the respective target polarization (corrected for depolarization effects).
In order to describe the Sivers SSA defined in 
Eq.~(\ref{Eq:AUT-Siv-unw-SIDIS-exp}) in Ref.~\cite{Collins:2005ie}
two major simplifications are made. First, soft factors 
\cite{Collins:1981uk,Ji:2004wu,Collins:2004nx} are neglected.
Second, for the distribution of transverse momenta in $D_1^a(z,K_T^2)$ 
and $f_{1T}^{\perp a}(x,p_T^2)$ the Gaussian model is assumed. 

The Gaussian model certainly oversimplifies the description of 
``unintegrated'' distribution or fragmentation functions which are an 
involved issue in QCD \cite{Collins:2003fm}. 
On a longer term, an approach to the $p_T$-dependence of the Sivers SSA along 
the lines of the formalism in Ref.~\cite{Collins:1984kg} would be desirable. 
The Gaussian model, however, provides a good effective description of 
SIDIS and DY data within a certain range of low transverse momenta, and
is sufficient for the purpose of the present work. The free parameters, 
namely the Gaussian widths, are consistently constrained in
\cite{Collins:2005ie} by the HERMES data. 

As regards the flavor dependence of the Sivers functions, there are no 
strong constraints from the present SIDIS data 
\cite{Airapetian:2004tw,Alexakhin:2005iw}.
In fact, in Ref.~\cite{Anselmino:2005nn} where this has been attempted all 
fitted distributions but $f_{1T}^{\perp u}$ were found consistent with 
zero. In this situation it is appealing to invoke additional theoretical 
constraints. In particular, here we use predictions from the 
QCD limit of a large number of colours $N_c$. 

\begin{figure}[t!]
{\epsfxsize=2in\epsfbox{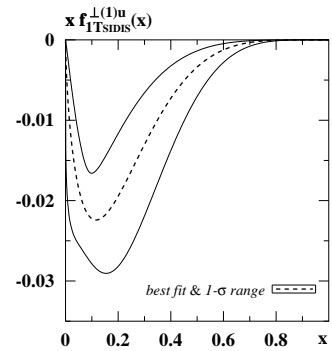}}
        \vspace{-0.3cm}
        \caption{\label{fig2-fit}\footnotesize\sl
        The $u$-quark Sivers function 
        $xf_{1T\SIDIS}^{\perp(1)u}(x)$ vs.\  $x$ at a scale of about 
        $2.5\,{\rm GeV}^2$, as obtained from a fit to the HERMES data 
        \cite{Airapetian:2004tw}.
        Shown are the best fit and its 1-$\sigma$ uncertainty.}
\end{figure}

In this limit the nucleon appears as $N_c$ quarks bound by a mean field
\cite{Witten:1979kh}, which exhibits certain spin-flavour symmetries 
\cite{Balachandran:1982cb}. By exploring these symmetry properties it 
was proven in a model independent way that in the large-$N_c$ limit
\cite{Pobylitsa:2003ty} 
\be\label{Eq:large-Nc}
      f_{1T}^{\perp u}(x,{\bf p}_T^2) =
    - f_{1T}^{\perp d}(x,{\bf p}_T^2) \;\;\;
    \mbox{modulo $1/N_c$ corrections,}\ee
for not too small and not too large $x$ satisfying $xN_c={\cal O}(N_c^0)$.
Analog relations hold for the Sivers antiquark distributions.\footnote{
        For historical correctness we mention that previously 
        (\ref{Eq:large-Nc}) was discussed in the framework of (simple 
        versions of) chiral models \cite{Anselmino:2001vn}. But
        the way in which (\ref{Eq:large-Nc}) was obtained there was 
        shown to be incorrect \cite{Pobylitsa:2002fr}.
        Recently, in Ref.~\cite{Drago:2005gz} (a more sophisticated
        version of) a chiral model with vector mesons obeying a hidden 
        local flavour symmetry was discussed, in which the Sivers 
        function would obey (\ref{Eq:large-Nc}).}
Imposing the large-$N_c$ relation (\ref{Eq:large-Nc}) as an additional
constraint, and neglecting effects of antiquarks and heavier flavours,
it is shown \cite{Collins:2005ie} that the HERMES data \cite{Airapetian:2004tw} 
can be described by the following 2-parameter ansatz and best fit
\ba\label{Eq:ansatz+fit} 
        f_{1T\SIDIS}^{\perp (1) u}(x) = -f_{1T\SIDIS}^{\perp (1) d}(x)
        &\stackrel{\rm ansatz}{=}&\;  A \, x^b   \,(1-x)^5\;, \\
        &\stackrel{\rm fit   }{=}&\!\!\!\! -0.17 x^{0.66}(1-x)^5\;, \nonumber
        \phantom{X}
        \ea
with a $\chi^2$ per degree of freedom of about $0.3$. The fit and 
its 1-$\sigma$ uncertainty due to the statistical error of the data 
\cite{Airapetian:2004tw} are shown in Fig.~\ref{fig2-fit}. Several comments 
are in order concerning the ansatz and fit result (\ref{Eq:ansatz+fit}).

The fit (\ref{Eq:ansatz+fit}) refers to a scale of $2.5\,{\rm GeV}^2$
roughly set by the $\la Q^2\ra$ in the HERMES experiment 
\cite{Airapetian:2004tw}. For the extraction we used the parameterizations 
for $f_1^a(x)$ and $D_1^a(x)$ from \cite{Gluck:1998xa,Kretzer:2001pz} at
the corresponding scale.

Within the Gaussian model, of course, one does not need to work with the
``transverse moment'' of the Sivers function defined as, and in the Gaussian 
ansatz given by
\ba\label{Eq:Def-Siv-transverse-mom}
        f_{1T}^{\perp(1)a}(x) 
        &\equiv& \int\!\di^2{\bf p}_T\; \frac{{\bf p}_T^2}{2 M_N^2}\;
                   f_{1T}^{\perp a}(x,{\bf p}_T^2) \\
        &\stackrel{\rm Gauss}{=}&
        \frac{\la p_T^2\ra_{\rm Siv}^{\phantom{2}}}{2 M_N^2}\;
        f_{1T}^{\perp a}(x) \;,
        \nonumber \ea
where $\la p_T^2\ra_{\rm Siv}^{\phantom{2}}$ denotes the Gaussian width of 
the Sivers function. However, doing so one benefits from the fact that the 
fit for the moment $f_{1T}^{\perp(1)a}(x)$ (in contrast to 
$f_{1T}^{\perp a}(x,{\bf p}_T^2)$) is  
nearly insensitive \cite{Collins:2005ie} to the value of 
$\la p_T^2\ra_{\rm Siv}^{\phantom{2}}$
which is poorly constrained by the data and the positivity bound 
\cite{Bacchetta:1999kz}.

The shape of the Sivers function at large $x$ is not constrained by
the data \cite{Airapetian:2004tw}.  Our ansatz of $(1-x)^5$ dependence
has some theoretical justification \cite{Efremov:2004tp}, although
there are also arguments \cite{Anselmino:2005ea} for a $(1-x)^4$
behavior.

We remark that the result (\ref{Eq:ansatz+fit}) is in good agreement with
the fit of Ref.~\cite{Efremov:2004tp} to the {\sl preliminary} HERMES data
\cite{HERMES-new} on the Sivers SSA weighted with a power of the transverse 
momentum $P_{h\perp}$ of the produced hadron. 
With such a weighting the SSA can be interpreted (neglecting soft factors) 
unambiguously in terms of $f_{1T}^{\perp(1)a}(x)$ independently of any 
model for transverse momentum dependence \cite{Boer:1997nt}.
Keeping in mind the {\sl preliminary} status of the data \cite{HERMES-new},
this agreement indicates the consistency of the Gaussian ansatz within the
accuracy of the data.

The large-$N_c$ motivated ansatz (\ref{Eq:ansatz+fit}) is confirmed 
by results from the COMPASS experiment \cite{Alexakhin:2005iw} where a 
solid polarized $^6$LiD target \cite{Ball:2003vb,Goertz:2002vv} was used. 
Neglecting nuclear binding effects and exploring isospin symmetry, in 
deuterium 
$f_{1T}^{u/D}\approx f_{1T}^{u/p}+f_{1T}^{u/n}\approx f_{1T}^u+f_{1T}^d$
and analogously for $d$, $\bar q$, etc.
Thus, the deuterium target is sensitive to the flavour combination
which is suppressed in the large-$N_c$ limit, see (\ref{Eq:large-Nc}), 
and for which our ansatz (\ref{Eq:ansatz+fit}) yields zero.
This is in agreement with the COMPASS data \cite{Alexakhin:2005iw}
showing a Sivers effect from deuterium target compatible with zero 
within error bars.

The reason why the large $N_c$ picture of the Sivers function works at the
present stage of art, is due to the fact that the current precision of the 
first data \cite{Airapetian:2004tw,Alexakhin:2005iw} is comparable to the 
theoretical accuracy of the large-$N_c$ relation (\ref{Eq:large-Nc}). 
Thus, $1/N_c$ corrections (and antiquarks effects) cannot be resolved within 
the error bars of the data \cite{Airapetian:2004tw,Alexakhin:2005iw}.
In future, when the precision of the data will increase, it will certainly
be necessary to refine the ansatz (\ref{Eq:ansatz+fit}) to include $1/N_c
$ corrections and antiquark effects.

The fit (\ref{Eq:ansatz+fit}) described above is to the final HERMES
data in Ref.\ \cite{Airapetian:2004tw}.  We observe that the fit is
compatible within the 1-$\sigma$ range with more precise, but {\sl
  preliminary}, data that has recently become available
\cite{Diefenthaler:2005gx}. Incorporating into our fit these data,
which are not yet corrected for smearing and acceptance effects, would
tend to increase somewhat the Sivers function, c.f.\ 
\cite{Anselmino:2005ea,Vogelsang:2005cs,Collins:2005wb,Anselmino:2005an}.

The sign of $f_{1T\SIDIS}^{\perp(1)u}(x)$ in Eq.~(\ref{Eq:ansatz+fit}) is
in agreement with the physical picture of the Sivers functions
discussed in Ref.~\cite{Burkardt:2002ks}.

We remark that, in principle, one could include into the fit in
addition to SIDIS data also the data on SSA in the hadronic
processes $p^\uparrow p \to\pi X$ or $p^\uparrow \bar p\to\pi X$
\cite{Adams:1991rw,Adams:2003fx}.  This possibility was not
explored in \cite{Collins:2005ie} as these SSA could also be due
to other mechanisms
\cite{Collins:1992kk,Efremov:eb,Kanazawa:2000hz}.  Furthermore, as
shown in \cite{Bomhof:2004aw} it is not clear that factorization
holds in these processes in terms of $p_T$-dependent correlators,
for reasons that do not apply to SIDIS and to the Drell-Yan
process.  One also has to keep in mind that it is a priori not
clear whether the leading twist factorization approach is accurate
at the $\la Q^2\ra=2.5\,{\rm GeV}^2$ of the HERMES experiment as
assumed in \cite{Collins:2005ie}. Only careful analyses (which
will include soft factors) of future data from experiments
performed at different $Q^2$ will reveal to what extent this
assumption is justified.

\section{\boldmath Sivers effect in SSA in DY at RHIC}
\label{Sec-3:Sivers-in-DY-at-RHIC}
         
The process $p^\uparrow p\to l^+l^-X$ is characterized by the variables 
$s=(p_1+p_2)^2$, the invariant mass of the lepton pair $Q^2=(k_1+k_2)^2$,
and the rapidity
%
%
\be
    y=\frac12 \, \ln\frac{p_2\cdot(k_1+k_2)}{p_1\cdot(k_1+k_2)} \;,
\ee
where $p_{1/2}$ (and $k_{1/2}$) indicate the momenta of the incoming proton
(and the outgoing lepton) pair.

Let us consider the azimuthal SSA defined as a sum over the events $i$ 
according to
\ba\label{Eq:DY-AUT-0}
&&  A_{UT}^{\sin(\phi-\phi_S)} \\
&&  = \frac{\sum_i\sin(\phi_i-\phi_{S,i})
    \{N^\uparrow(\phi_i,\phi_{S,i})-N^\downarrow(\phi_i,\phi_{S,i}+\pi)\}}
    {\frac12\sum_i
    \{N^\uparrow(\phi_i,\phi_{S,i})+N^\downarrow(\phi_i,\phi_{S,i}+\pi)\}} \;,
        \nonumber\ea
where $\uparrow,\downarrow$ denote the transverse polarizations of the proton,
the polarized proton moves in the positive $z$-direction, and $(\phi-\phi_S)$ is
the azimuthal angle between the virtual photon and the polarization vector. 
Neglecting again soft factors and assuming the Gaussian model for the 
distribution of transverse momenta, to leading order the SSA is given by
\be\label{Eq:DY-AUT-1}
    A_{UT}^{\sin(\phi-\phi_S)} = 2\;\frac{a_{\rm Gauss}^\DY
    \sum_a e_a^2 \, x_1f_{1T\,\DY}^{\perp(1) a}(x_1)\,x_2f_1^{\bar a}(x_2)}{
    \sum_a e_a^2 \, x_1f_1^{a}                 (x_1)\,x_2f_1^{\bar a}(x_2)}\;,
    \ee
where $a=u,\bar{u},d,\bar{d}$, etc.\  and the parton momenta $x_{1,2}$ are 
given by $x_{1,2}=(Q^2/s)^{1/2}\,e^{\pm y}$. The dependence on the Gaussian 
model is contained in the factor
\be\label{Eq:DY-AUT-1a}
        a_{\rm Gauss}^\DY = \frac{\sqrt{\pi}}{2}\,
        \frac{M_N}{\sqrt{\la p_T^2\ra_{\rm Siv}^{\phantom{2}}
                        +\la p_T^2\ra_{\rm unp}^{\phantom{2}}}}\;.
\ee
If one introduced in the numerator of (\ref{Eq:DY-AUT-0}) an additional weight
$q_T/M_N$, where $q_T$ denotes the modulus of the transverse momentum of 
the lepton pair with respect to the collision axis, the resulting SSA would 
be independent of a particular model for transverse momentum distribution 
and given by the expression on the right-hand-side of (\ref{Eq:DY-AUT-1}) 
but without the factor $a_{\rm Gauss}^\DY$. 
It was argued that such a ``transverse momentum weighted'' SSA might be
protected against Sudakov dilution effects \cite{Boer:2001he}. 
These effects need not be negligible, considering the fact that 
at RHIC we deal with SSA at considerably higher scales than in the
HERMES experiment: $(4\,{\rm GeV})^2 \leq Q^2 \leq (20\,{\rm GeV})^2$
typically at RHIC \cite{Bunce:2000uv} vs.\  $\la Q^2\ra = 2.5\,{\rm GeV}^2$ 
at HERMES.

%
\begin{figure*}[t!]
\begin{tabular}{cc}
    {\epsfxsize=2.4in\epsfbox{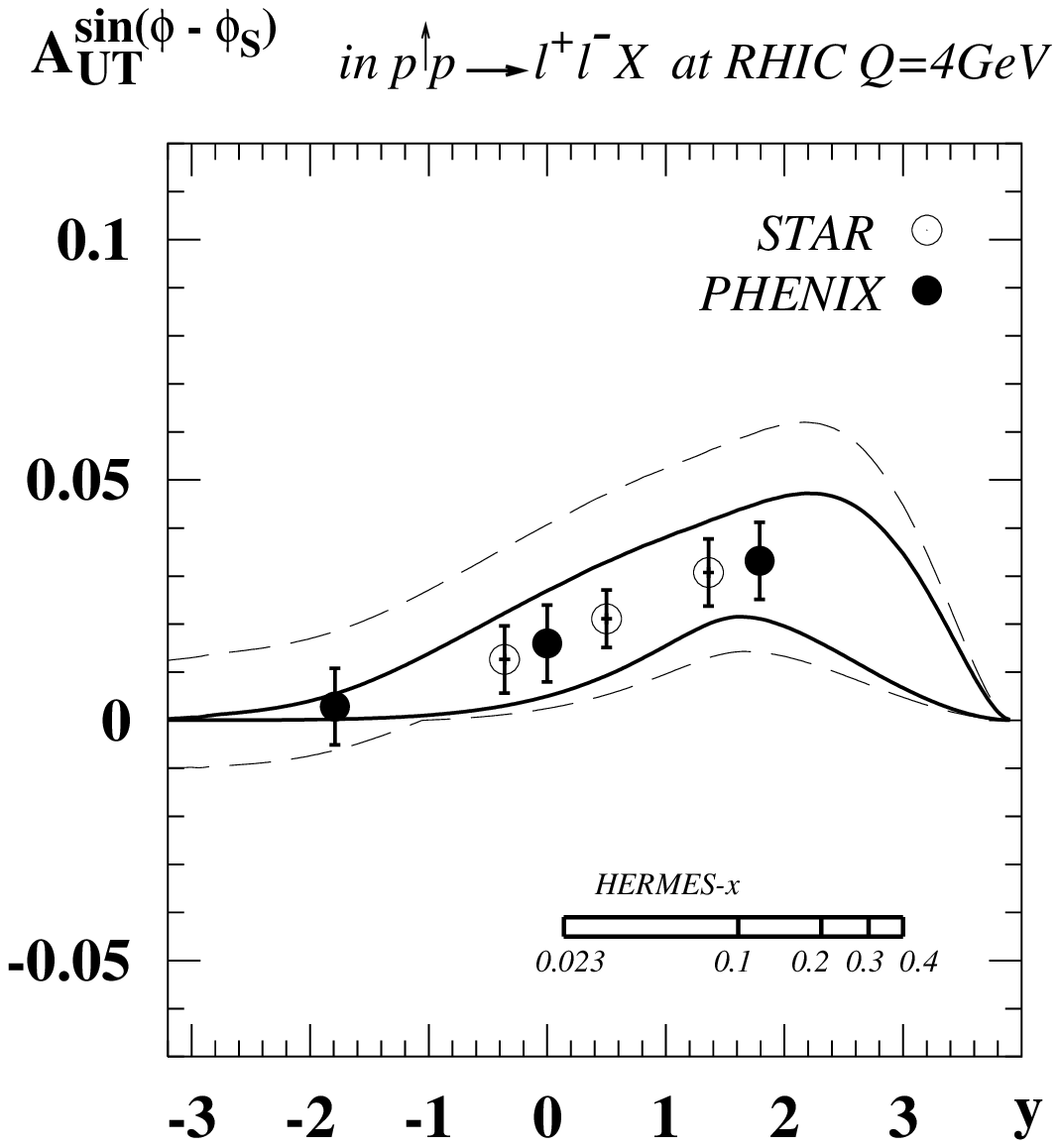}} &
    {\epsfxsize=2.4in\epsfbox{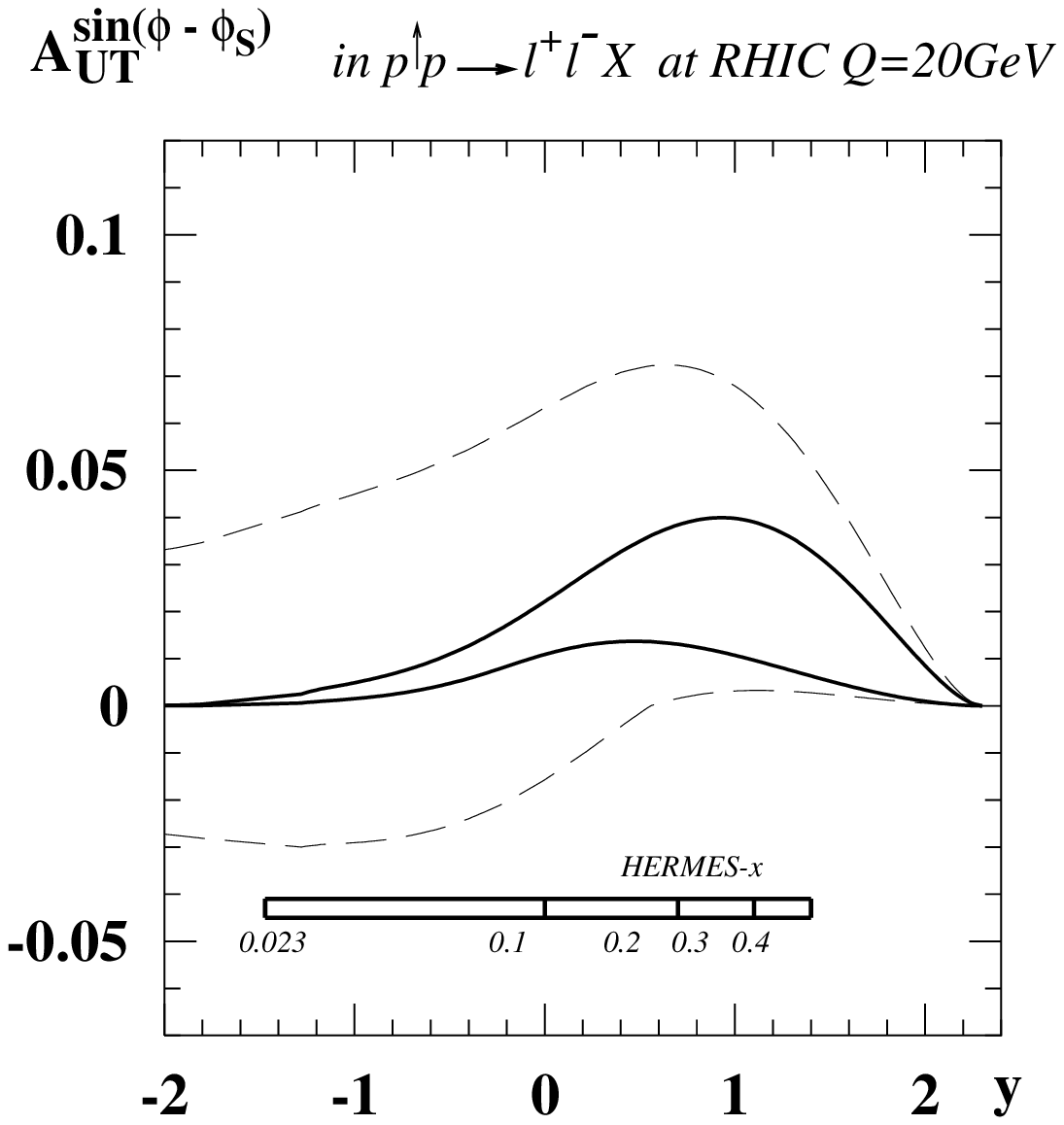}} \cr
    {\epsfxsize=2.4in\epsfbox{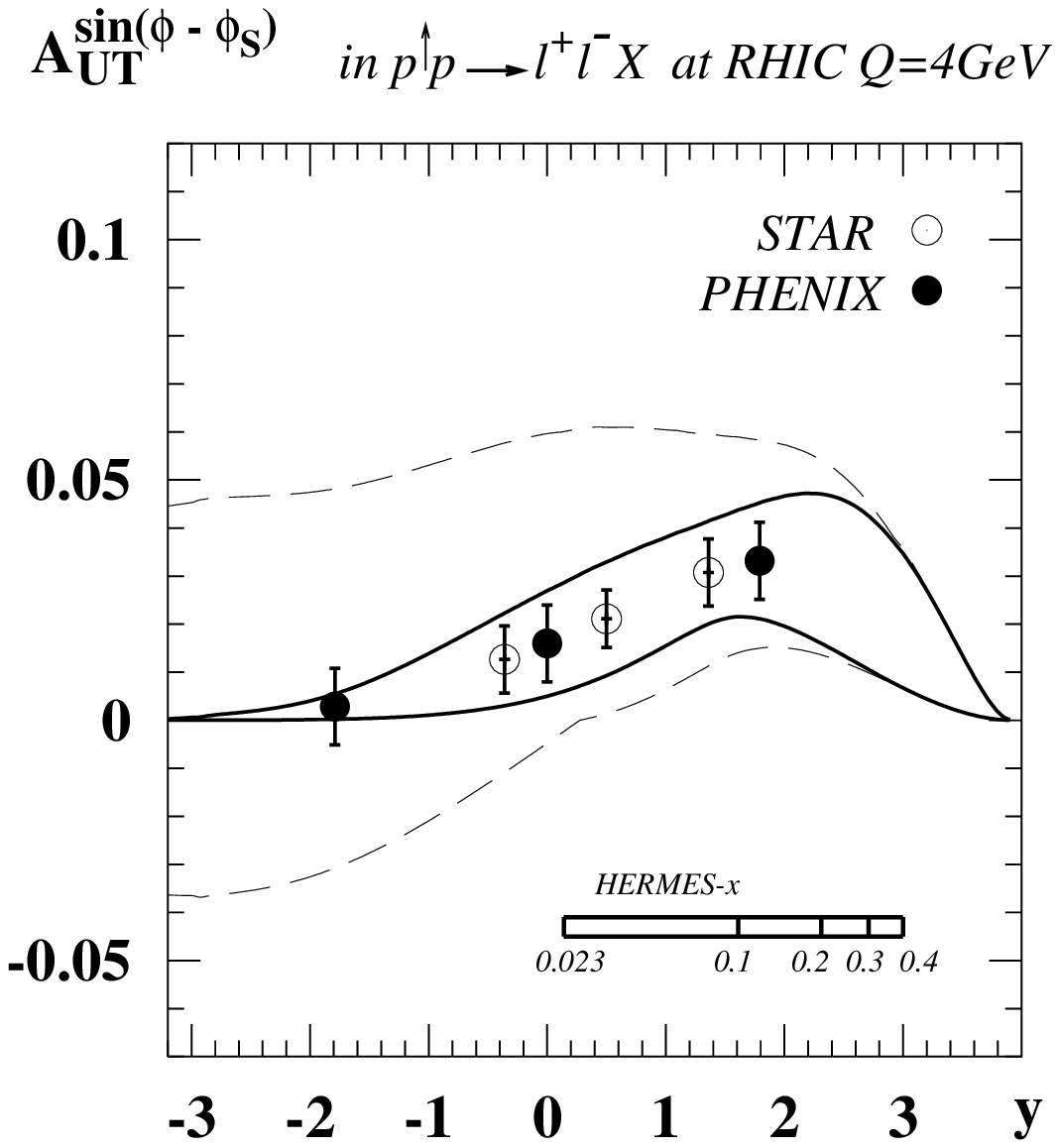}} &
    {\epsfxsize=2.4in\epsfbox{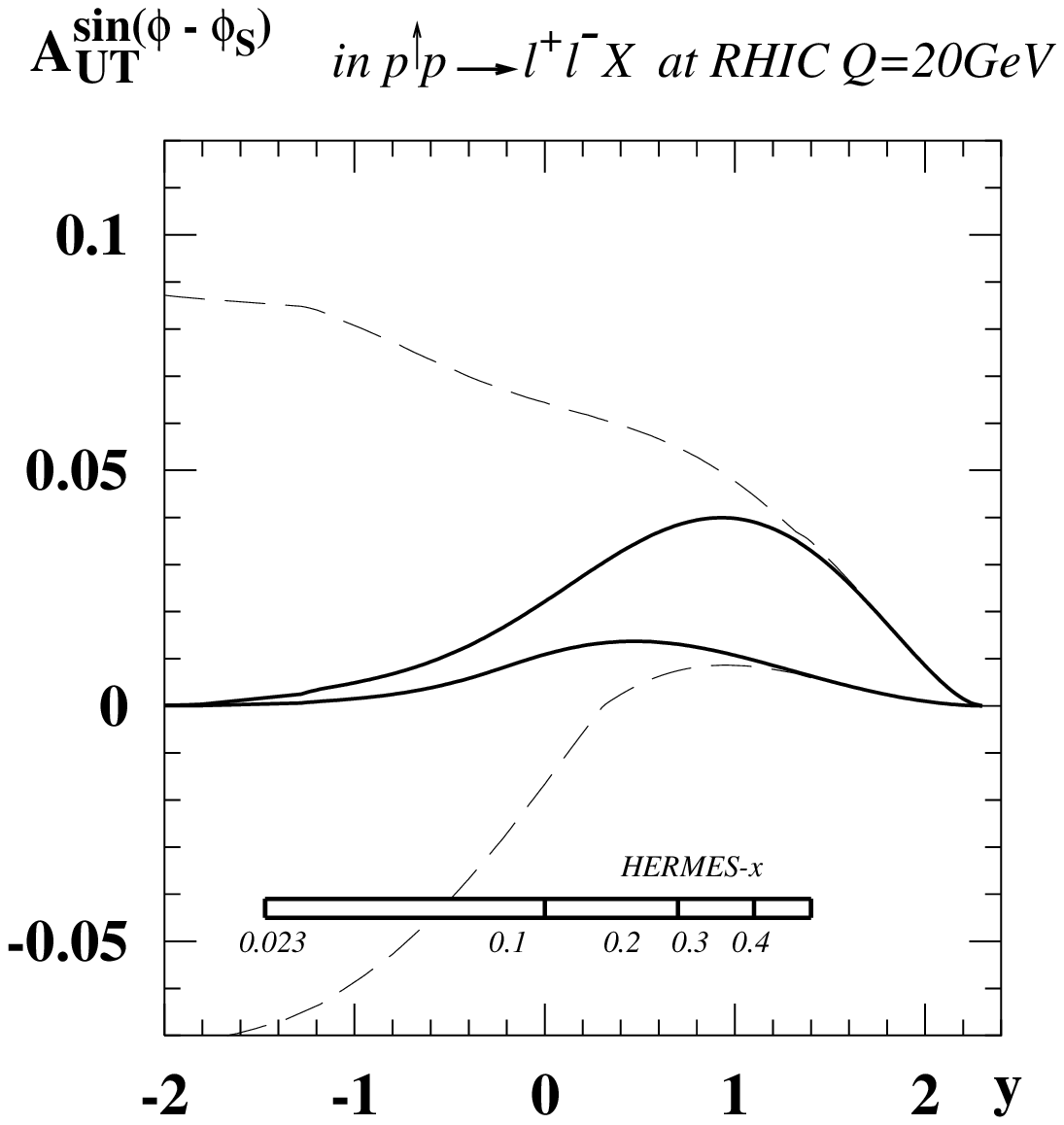}} \cr 
\end{tabular}
\caption{\footnotesize\sl
    \label{Fig2-DY-at-RHIC} 
    The azimuthal SSA $A_{UT}^{\sin(\phi-\phi_S)}$ in Drell-Yan lepton
    pair production, $p^\uparrow p\to l^+l^- X$, as function of $y$ for the
    kinematics of the RHIC experiment with $\sqrt{s}=200\,{\rm GeV}$.
    The left plots show $Q^2=(4\,{\rm GeV})^2$, and the right plots
    $Q^2=(20\,{\rm GeV})^2$.  The upper plots show the effects of
    varying the Sivers antiquark distributions within the range of
    model I --- see
    Eqs.~(\ref{Eq:model-Sivers-qbar},~\ref{Eq:model-Sivers-qbar2}).
    The lower plots show instead the results of model II.  In all
    cases, the central estimates are based on our fit
    (\ref{Eq:ansatz+fit}) for the Sivers quark distribution functions
    from the HERMES data \cite{Airapetian:2004tw}.  The inner error
    band (solid lines) shows the 1-$\sigma$ uncertainty of the fit. The
    outer error bands (dashed lines) show the error from varying the Sivers 
    antiquark distribution functions within the ranges specified in
    Eqs.~(\ref{Eq:model-Sivers-qbar},~\ref{Eq:model-Sivers-qbar2}), 
    model I for the upper plots, and model II for the lower plots.
    The $x$-region explored in the HERMES kinematics is indicated.
    For $Q=4\,{\rm GeV}$ we show the estimated
    statistical error for STAR and PHENIX. For $Q=20\,{\rm GeV}$ the 
    statistical error at STAR and PHENIX is comparable to the asymmetry.  
    See text and Table~\ref{Table-with-error-estimates} for a discussion 
    of the planned RHIC II.}
\end{figure*}
%

The DY process in $pp$-collisions is sensitive to 
$f_{1T\,\DY}^{\perp(1) q}(x_1)f_1^{\bar q}(x_2)$ and to
$f_{1T\,\DY}^{\perp(1) \bar q}(x_1)f_1^{q}(x_2)$ on an equal footing.
Thus, even though the effects of the Sivers antiquark distributions
seem small and presently not observable in SIDIS, one cannot expect
this to be the case in DY at RHIC.

In order to have a rough idea which regions of $y$ (for a given $Q^2$) could 
be more and which regions less sensitive to Sivers antiquark distribution
functions, let us introduce two {\sl models}:
\be\label{Eq:model-Sivers-qbar}
        f_{1T\DY}^{\perp(1) \bar q}(x) = \epsilon(x) \;
        f_{1T\DY}^{\perp(1)      q}(x) \;,
\ee
with
\be\label{Eq:model-Sivers-qbar2}
        \epsilon(x) = \pm\,\cases{
        0.25 = {\rm const} & model I \cr & \cr
        \displaystyle\frac{(f_1^{\bar u}+f_1^{\bar d})(x)}{(f_1^u+f_1^d)(x)}
        & model II}
\ee
where for $f_{1T\,\DY}^{\perp(1) q}(x)$ we will use the result from 
Eq.~(\ref{Eq:ansatz+fit}) taking into account the change of sign in 
Eq.~(\ref{Eq:01}).

Model I is that the Sivers  antiquark distribution is $\pm 25\%$ 
of the corresponding Sivers quark distribution at all $x$. 
Model II is that the ratio of the Sivers antiquark over quark distribution 
is fixed by the ratio of the unpolarized distribution functions.
The particular ansatz for model II ensures compatibility with the 
large-$N_c$ limit, and it preserves the following sum rule
\cite{Burkardt:2003yg} --- see also \cite{Efremov:2004tp} ---
\be\label{Eq:th-05}
    \sum_{a=g,u,d,\,\dots}
    \int\!\di x \;f_{1T}^{\perp(1)a}(x) = 0\;.
\ee
(Note that in the large-$N_c$ limit the gluon Sivers distribution
is suppressed with respect to the quark one \cite{Efremov:2004tp}.)

Eqs.~(\ref{Eq:model-Sivers-qbar},~\ref{Eq:model-Sivers-qbar2}) represent 
rough models for $f_{1T\,\DY}^{\perp(1) \bar q}(x)$ which are, however, 
consistent with the HERMES data \cite{Airapetian:2004tw}, and
compatible with the presently known theoretical constraints, namely
the large-$N_c$ relations (\ref{Eq:large-Nc}), positivity constraints 
\cite{Bacchetta:1999kz} and the Burkardt-sum rule (\ref{Eq:th-05}).
This makes them sufficient for our purposes. 

In order to present estimates for RHIC we strictly speaking should use
the fitted Sivers function (and $f_1^a(x)$) evolved to the relevant
scale. Instead, let us assume that 
\be\label{Eq:neglect-scale-dependence}
        \frac{\sum_ae_a^2f_{1T\,\DY}^{\perp(1)a}f_1^{\bar a}}
             {\sum_ae_a^2f_1^a f_1^{\bar a}}
        \biggl|_{{\rm HERMES}\,Q^2} \approx 
        \frac{\sum_ae_a^2f_{1T\,\DY}^{\perp(1)a}f_1^{\bar a}}
             {\sum_ae_a^2f_1^a f_1^{\bar a}}
        \biggl|_{{\rm RHIC}\,Q^2} \;.
\ee
It is difficult to quantify exactly the error introduced in this way.
However, we believe it to be smaller than other uncertainties in our analysis.

When dealing with unintegrated distribution functions at higher energies one 
must take into account a broadening of the average transverse momentum.
Considering that this effect due to radiating off soft gluons does not 
depend on polarization, we roughly estimate
\ba
&&               \la p_T^2\ra_{\rm Siv}^{\rm RHIC} 
        \approx 2\la p_T^2\ra_{\rm Siv}^{\rm HERMES}\;,\nonumber\\
&&               \la p_T^2\ra_{\rm unp}^{\rm RHIC} 
        \approx 2\la p_T^2\ra_{\rm unp}^{\rm HERMES} 
        \approx 0.66\,{\rm GeV}^2\;. \label{Eq:broadening} \ea
Similar values for $\la p_T^2\ra_{\rm unp}$ at high energies 
were obtained in \cite{D'Alesio:2004up}.
Next, let us assume that the transverse moment $f_{1T\,\DY}^{\perp(1)a}$ 
is little affected by that. 
The combined effect of our assumptions is that the SSA decreases
at higher energies -- in qualitative agreement with \cite{Boer:2001he},
though not as fast as discussed there.

We recall, that from the HERMES data \cite{Airapetian:2004tw} and the 
positivity bound \cite{Bacchetta:1999kz} the Gaussian width of the Sivers
function is only poorly constrained. Under the assumption (\ref{Eq:broadening})
one may expect it at RHIC to be in the range 
$\la p_T^2\ra_{\rm Siv}^{\rm RHIC}\approx(0.20\dots0.64)\,{\rm GeV}^2$
\cite{Collins:2005ie}. 
However, the only place where this value numerically matters is the 
Gauss factor in Eq.~(\ref{Eq:DY-AUT-1a}).
Varying $\la p_T^2\ra_{\rm Siv}^{\rm RHIC}$ in the above range yields
$a_{\rm Gauss}^\DY= 0.81 \cdot (1\pm 10\%)$, 
i.e. it alters our estimates for RHIC only within $\pm 10\%$.

On the basis of the above assumptions and taking into account the change
of sign for the Sivers distribution function in Eq.~(\ref{Eq:01}) we obtain 
the results shown in Fig.~\ref{Fig2-DY-at-RHIC}. We observe that the 
Drell-Yan SSA is noticeably sensitive to the Sivers $\bar q$-distribution 
functions, as modeled in Eq.~(\ref{Eq:model-Sivers-qbar}).
The effect is more pronounced at larger dilepton masses $Q$.
However, we note that at $Q=4\,{\rm GeV}$ in the region of positive 
rapidities $y$ the effect is moderate, and {\sl does not} alter the sign 
of the asymmetry. 

The region of negative rapidities $y$ is strongly sensitive to the Sivers 
$\bar q$-distribution. For positive $\epsilon(x)$ in 
Eqs.~(\ref{Eq:model-Sivers-qbar},~\ref{Eq:model-Sivers-qbar2}) the SSA is 
positive, for negative $\epsilon(x)$ it is negative. As already mentioned,
the effect of Sivers-$\bar q$ is more pronounced at larger $Q^2$. 
Thus, by focusing on these regions of $y,\,Q^2$ one could gain the first
information on the magnitude and sign of the Sivers-$\bar{q}$ distributions.

In order to see explicitly in which kinematical region our predictions are 
constrained by data, we show in Fig.~\ref{Fig2-DY-at-RHIC} the
$x$-range covered 
in the HERMES experiment which constrained the fit of the Sivers function. 
In this context it is worthwhile remarking that our estimates for RHIC 
based on SIDIS data are complementary to those made in Ref.~\cite{Anselmino:2002pd}.
There information on the Sivers function was used from the data on SSA in 
$p^\uparrow p\to \pi X$ \cite{Adams:1991rw}. 
Assuming factorization for this process and considering that other mechanisms cannot 
explain the observed SSA at large $x_F$ \cite{Anselmino:2004ky,Ma:2004tr}, one finds
that the data \cite{Adams:1991rw} constrain the Sivers function at large $x>0.4$ 
above the region explored at HERMES. 
Thus, in the region of large positive $y$, where our estimates are not 
constrained by SIDIS data, see Fig.~\ref{Fig2-DY-at-RHIC}, the estimates of 
Ref.~\cite{Anselmino:2002pd} could be more reliable.

%
\begingroup
\squeezetable
\begin{table}[h!]
\begin{ruledtabular}
    \begin{tabular}{l | ccc | ccc | ccc}
&&&&&&&&&\\
&\multicolumn{3}{ c|}{STAR} &
 \multicolumn{3}{ c|}{PHENIX} & 
 \multicolumn{3}{ c }{RHIC II} \\
&&&&&&&&&\\
\hline
&&&&&&&&&\\
 & $y$ & $4\,{\rm GeV}$ & $20\,{\rm GeV}$ 
 & $y$ & $4\,{\rm GeV}$ & $20\,{\rm GeV}$ 
 & $y$ & $4\,{\rm GeV}$ & $20\,{\rm GeV}$ \\
& -0.5 & 0.007 & 0.09 & -1.8 & 0.008 & 0.2 & $\pm 2.5$ & 0.003 & 0.03 \\
$\delta A$ 
&  0.5 & 0.006 & 0.06 &  0.0 & 0.017 & 0.13 & $\pm 1.5$ & 0.001 & 0.01 \\
&  1.5 & 0.007 & 0.11 &  1.8 & 0.008 & 0.2 & $\pm 0.5$ & 0.001 & 0.01 \\
&&&&&&&&&\\
\hline
&&&&&&&&&\\
$\int\!L\di t$ &
\multicolumn{3}{ c|}{$125\,{\rm pb}^{-1}$} &
\multicolumn{3}{ c|}{$125\,{\rm pb}^{-1}$} & 
\multicolumn{3}{ c }{$10\times125\,{\rm pb}^{-1}$}\\
&&&&&&&&&
\end{tabular}
\end{ruledtabular}
        \caption{\label{Table-with-error-estimates}\footnotesize\sl 
       Statistical errors $\delta A$ for the Sivers SSA in Drell Yan for the 
       PHENIX and STAR detectors at RHIC: Errors are shown for dilepton masses    
       of $Q=4$\,GeV and $20$\, GeV assuming an integrated luminosity of
       $\int Ldt = 125 pb^{-1}$ and a beam polarization of $P=0.7$. Error   
       estimates have been carried out using the event generator PYTHIA. 
       Projected errors are also shown for a possible future dedicated  
       experiment for transverse spin with large acceptance at 
       RHIC II (luminosity upgrade); see text for details.}
\end{table}
\endgroup
%

In Table~\ref{Table-with-error-estimates} we show the statistical errors 
$\delta A$ for the Sivers SSA in DY estimated with PYTHIA considering the 
acceptance of the STAR and PHENIX detectors. 
Detector acceptance is conveniently specified in terms of
pseudo-rapidity $\eta={\rm ln}\,({\rm tan}\frac{\theta}{2})$, which is directly 
related to the scattering angle $\theta$ of the lepton pair with respect to the 
beam-axis, and thus to the geometry of the detector. In the following acceptance 
cuts imposed on leptons will be given in pseudo rapditity. However,
asymmetries and their errors are analyzed in bins of photon rapidity $y$.
We assume an integrated 
luminosity $\int\! L\di t=125\,{\rm pb}^{-1}$ and a beam polarization of 
$P=0.7$. We use these parameters as an upper estimate for the statistics 
these experiments could acquire with transverse beam polarization before
RHIC detector and luminosity upgrades will become available in 2012 
(RHIC II). The statistical errors can be easily scaled to different 
parameters for integrated luminosity and beam polarization.

For STAR, which covers the range $ -1 < \eta < 2$ for $e^+e^-$, 
estimates for $\delta A$ are presented for bins centered at $y = -0.5$, 
$0.5$ and $1.5$. 
For PHENIX, which covers $|\eta|<0.35$ (for $e^+e^-$) and 
$1.2<|\eta|<2.4$ (for $\mu^+\mu^-$) we have chosen bins centered at 
$y=0$ and $|y|=1.8$. The bins in the dilepton mass $Q$ are chosen 
respectively at $4\,{\rm GeV}$ and $20\,{\rm GeV}$.

For $Q=4\,{\rm GeV}$ the Sivers SSA can be measured at STAR and PHENIX. For 
illustrative purposes the estimated $\delta A$ for STAR and PHENIX are shown 
in Fig.~\ref{Fig2-DY-at-RHIC}. The $\delta A$ at $Q=20\,{\rm GeV}$ is of the 
order of magnitude of the asymmetry itself.

The region of higher $Q$ could, however, be addressed taking advantage of the
higher luminosity available at RHIC II. We consider a new large acceptance
experiment dedicated to Drell Yan physics with transverse spin. The new detector could be located at one of the RHIC interaction regions which are not equipped with spin rotator magnets and therefore always provide proton-proton collisions
with transverse beam polarization.
 
We assume that at RHIC II the luminosities will be higher by a factor $2.5$ 
through electron cooling and that an additional factor $4$ in luminosity will 
result for the new experiment from special focusing magnets close to the interaction region. We expect that for a multi-year run with the new RHIC II 
detector an integrated luminosity of $10\times 125\,{\rm pb}^{-1}$ becomes 
easily accessible. The proposed detector covers an acceptance of 
$|\eta|<3.0$. In Table~\ref{Table-with-error-estimates} we present estimates 
for $\delta A$ for bins centered at $|y|=2.5$, $1.5$, $0.5$. 
Clearly, RHIC II could access the Sivers effect also at large dilepton 
masses, where the effects of Sivers antiquarks are expected to be
more pronounced.

A ``symmetric'' (with respect to $\eta$) detector will allow to gain a 
factor $2$ in statistics for SSA compared to double spin asymmetries, as 
either beam polarization can be used. This additional factor is not reflected
in the estimates presented in Table II.

\section{\boldmath $q_T$-dependence of SSA}
\label{Sec-4:qT-depencence-in-DY-at-RHIC}

In the previous Sections we have discussed SSA integrated over 
transverse  momenta, namely over $q_T$ of the lepton pair in the DY 
process (or, $P_{h\perp}$ of the produced hadron in SIDIS). However, 
SSA as functions of $q_T$ are equally interesting observables.

%
\begin{figure*}[t!]
\begin{tabular}{cc}
    {\epsfxsize=2.4in\epsfbox{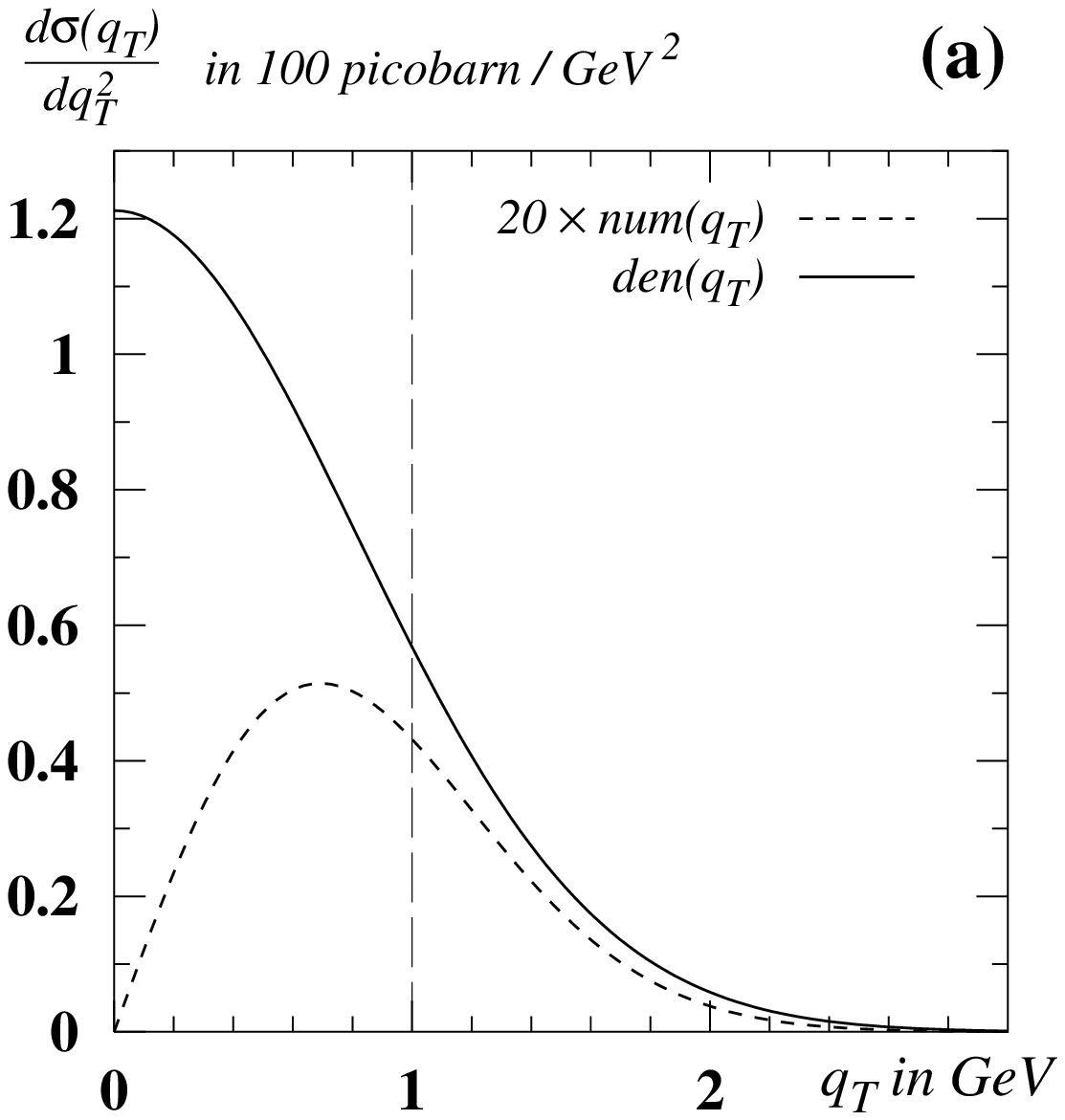}} &
    {\epsfxsize=2.4in\epsfbox{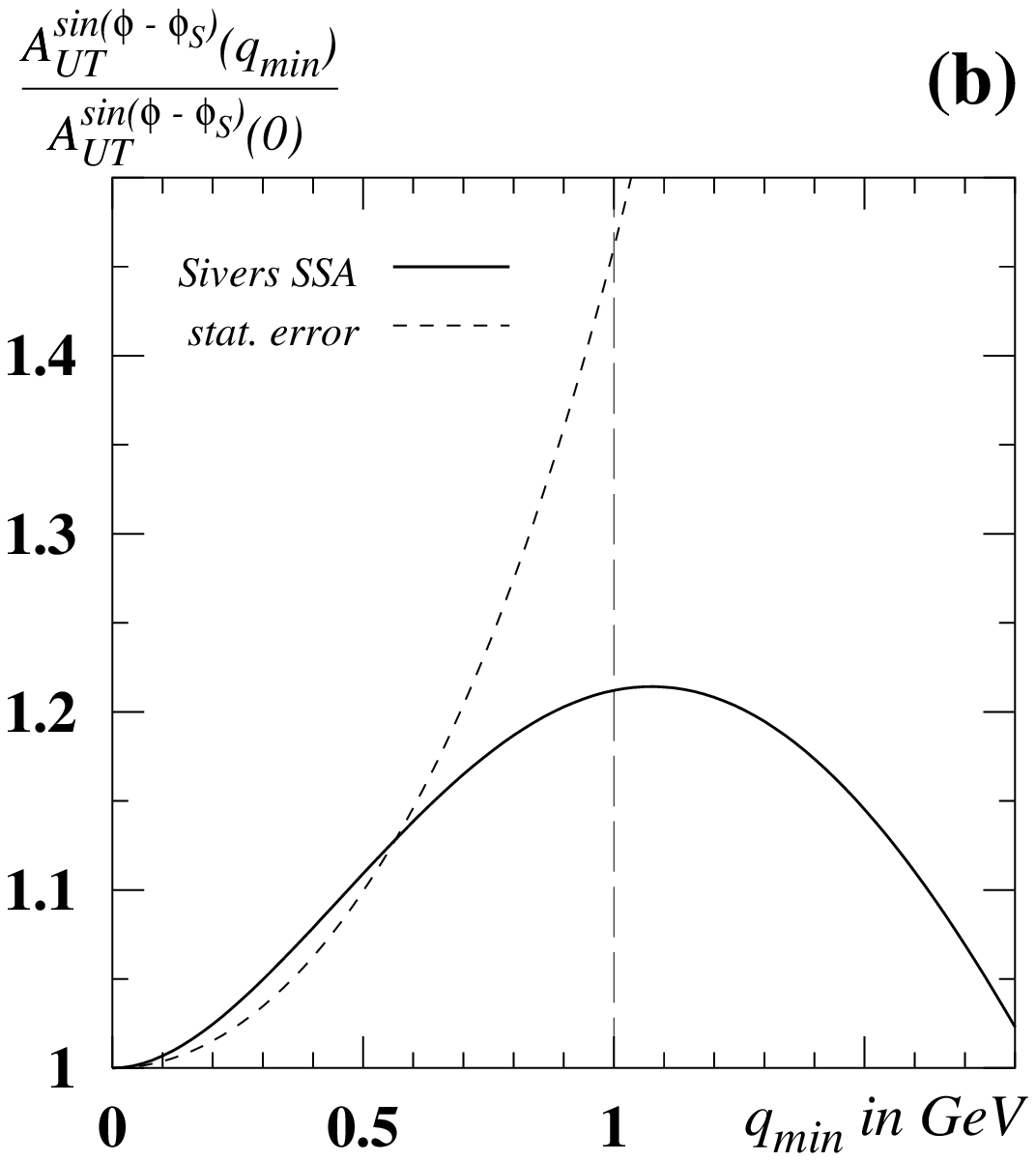}}
\end{tabular}
\caption{\footnotesize\sl
    \label{Fig3-DY-at-RHIC-qT} 
    {\bf a.} The numerator and denominator, see
    Eqs.~(\ref{AUT-DY-qT-0}--\ref{AUT-DY-qT-2}), for the azimuthal SSA 
    $A_{UT}^{\sin(\phi-\phi_S)}$ in the DY process as functions of the dilepton 
    transverse momentum $q_T$ for 
    $1.2<y<2.4$ at RHIC with $\sqrt{s}=200\,{\rm GeV}$ and 
    $Q^2=(4\,{\rm GeV})^2$. The absolute numbers for the cross sections are 
    somewhat altered by the assumption (\ref{Eq:neglect-scale-dependence}).
    {\bf b.} 
    The azimuthal SSA $A_{UT}^{\sin(\phi-\phi_S)}$ and its estimated
    statistical uncertainty as functions of the low-$q_T$ cut $q_{min}$,
    i.e.\  the ratio of the numerator and denominator in 
    Fig.~\ref{Fig3-DY-at-RHIC-qT}a integrated respectively over
    $q_T\in[q_{min},\infty]$. Both the SSA and its uncertainty are
    normalized with respect to their values for $q_{min}=0$.}
\end{figure*}
%

In the case of the DY Sivers SSA (\ref{Eq:DY-AUT-1}) we were able to 
minimize the model dependence to a certain extent (within the Gaussian 
model), e.g., by showing that the estimated SSA varied only within $\pm10\%$ 
when the Gaussian width of the Sivers function was allowed to vary over a wide 
range  $\la p_T^2\ra_{\rm Siv}^{\rm RHIC}\approx(0.20\dots0.64)\,{\rm GeV}^2$. 
When discussing the DY Sivers SSA as a function of $q_T$ our results are 
model-dependent. Nevertheless certain 
features are of a general character and it is worthwhile addressing 
them here. In what follows we assume for illustrative purposes that
$\la p_T^2\ra_{\rm Siv}^{\rm RHIC}\approx 0.3\,{\rm GeV}^2$. 

It is worthwhile stressing that unpolarized DY data on
the $q_T$ dependence can be well described by means of the Gaussian
model up to $q_T\lesssim (2$-$3)\,{\rm GeV}$ (depending to some extent on 
the invariant mass $Q$ and the kinematics of the respective experiment, see 
the detailed study in Ref.~\cite{D'Alesio:2004up}).
(For a description of the $q_T$ dependence of unpolarized DY in terms of 
the QCD-based formalism of \cite{Collins:1984kg} we refer to the 
work \cite{Landry:2002ix} and references therein.)
It is by no means clear whether the transverse parton momentum dependence
of the Sivers function can be described equally satisfactory by the Gauss 
ansatz. This, in fact, is among what we are going to learn from RHIC and 
other experiments.

Let us first mention an unrealistic property of the Gauss model.
Eq.~(\ref{Eq:DY-AUT-1}) was obtained upon integrating the relevant transverse 
momentum dependent cross sections over $q_T$ from 0 to $\infty$. In QCD the 
corresponding diagrams diverge, and physically it is clear that the large 
$q_T$ must be cutoff at a scale $\sim{\cal O}(Q)$  \cite{Collins:2003fm}. 

Experimentally no artificial large-$q_T$ cutoff is needed, since
correct kinematics imposes a kinematic limit. What is more interesting
in our context is to 
impose a low-$q_T$ cutoff. This makes sense, and could be even preferable, 
because it could allow to increase the asymmetry, for the Sivers SSA 
has a kinematical zero, i.e.\  it is SSA$\,\propto q_T$.

Let us define the $q_T$-dependent DY Sivers SSA as
\be\label{AUT-DY-qT-0}
    A_{UT}^{\sin(\phi-\phi_S)}(q_T) = \frac{{\rm num}(q_T)}{{\rm den}(q_T)}
\ee
with the numerator and denominator given by
\ba\label{AUT-DY-qT-1}
        {\rm num}(q_T) 
        &=& 
        \frac{\sigma_{UT}^{\rm Siv}\;M_N}
             {(\la p_T^2\ra_{\rm Siv}^{\rm RHIC}
              +\la p_T^2\ra_{\rm unp}^{\rm RHIC})^2}\,\nonumber\\
&&      \times \;q_T\; \exp\left(-\,
        \frac{q_T^2}{\la p_T^2\ra_{\rm Siv}^{\rm RHIC}
                    +\la p_T^2\ra_{\rm unp}^{\rm RHIC}}\right)\;,
        \nonumber\\
        {\rm den}(q_T) &=& 
        \frac{\sigma_{UU}}{2\la p_T^2\ra_{\rm unp}^{\rm RHIC}}
        \exp\left(-\,\frac{q_T^2}{2\la p_T^2\ra_{\rm unp}^{\rm RHIC}}\right)\;,
        \ea
where
\ba\label{AUT-DY-qT-2}
        \sigma_{UT}^{\rm Siv}
        &=& \int\!\!\!\!\int\limits_{\!\!\!\!\!\rm cuts}\di Q^2\di y
            \frac{4\pi\alpha^2}{9Q^4}\sum_ae_a^2
            x_1f_{1T\,\DY}^{\perp(1)a}(x_1)\,x_2f_1^{\bar a}(x_2)\;,\nonumber\\
        \sigma_{UU}
        &=&
        \int\!\!\!\!\int\limits_{\!\!\!\!\!\rm cuts}\di Q^2\di y
        \frac{4\pi\alpha^2}{9Q^4}\sum_ae_a^2 
        x_1f_1^{a}(x_1)\,x_2f_1^{\bar a}(x_2)\;.
\ea
Integrating ${\rm den}$ (or ${\rm num}$) over $q_T^2$ from 0 to 
$\infty$ yields the total unpolarized DY cross section $\sigma_{UU}$
(or $a_{\rm Gauss}^\DY \sigma_{UT}^{\rm Siv}$).

Considering a polarization-independent $K_{\rm factor}\approx 1.5$ and under 
the above discussed assumptions we obtain for the (quark dominated) kinematics
in the forward region $1.2<y<2.4$ for $Q=(4$-$5)\,{\rm GeV}$ the results shown 
in Fig.~\ref{Fig3-DY-at-RHIC-qT}a. 
(The uncertainty of ${\rm num}(q_T)$ due to the 1-$\sigma$ region of the 
HERMES fit and the effect of antiquarks is:
$K_{\rm factor}\sigma_{UT}^{\rm Siv}=(6.0\pm 3.5)\cdot10^{-12}\,{\rm barn}$.
For better visibility Fig.~\ref{Fig3-DY-at-RHIC-qT}a shows only the central 
value.)

Clearly, by including the low-$q_T$ region the SSA is diminished.
One could increase the SSA by applying a low-$q_T$ cut. The effect
of such a cut is illustrated in Fig.~\ref{Fig3-DY-at-RHIC-qT}b.
E.g., choosing $q_{min}=1\,{\rm GeV}^2$ increases the SSA by about 
$20\%$ which is close to the optimum for our choice of parameters.\footnote{
        \label{Footnote-choose-diiefferent-parameter}
        Had we chosen $\la p_T^2\ra_{\rm Siv}^{\rm RHIC}=0.6\,{\rm GeV}^2$ the 
        gain would be much more spectacular, namely up to a factor of 2.5.
        The optimum would then, however, appear for $q_{min}=3\,{\rm GeV}^2$,
        i.e. beyond the presumed range of applicability of the Gauss model.
        Also the loss of statistics would then be considerable.}
Of course, there is a price to pay for. The statistical error 
$\delta A_{UT}^{\sin(\phi-\phi_S)}\propto 1/\sqrt{N}$ grows by $\sim50\%$
due to loss of statistics, i.e.\ fewer counts $N$. In principle, one can
find a favoured $q_{min}$, which would allow to achieve the
maximal relative accuracy in the experiment.
This is a good illustration of the old ``golden rule'' of spin physics:
the best analyzing power is in the kinematical region where 
$(\mbox{spin asymmetry})^2\times(\mbox{number of counts})$ reaches a maximal value.

%
\begin{figure*}[ht!]
\begin{tabular}{cc}
    {\epsfxsize=2.4in\epsfbox{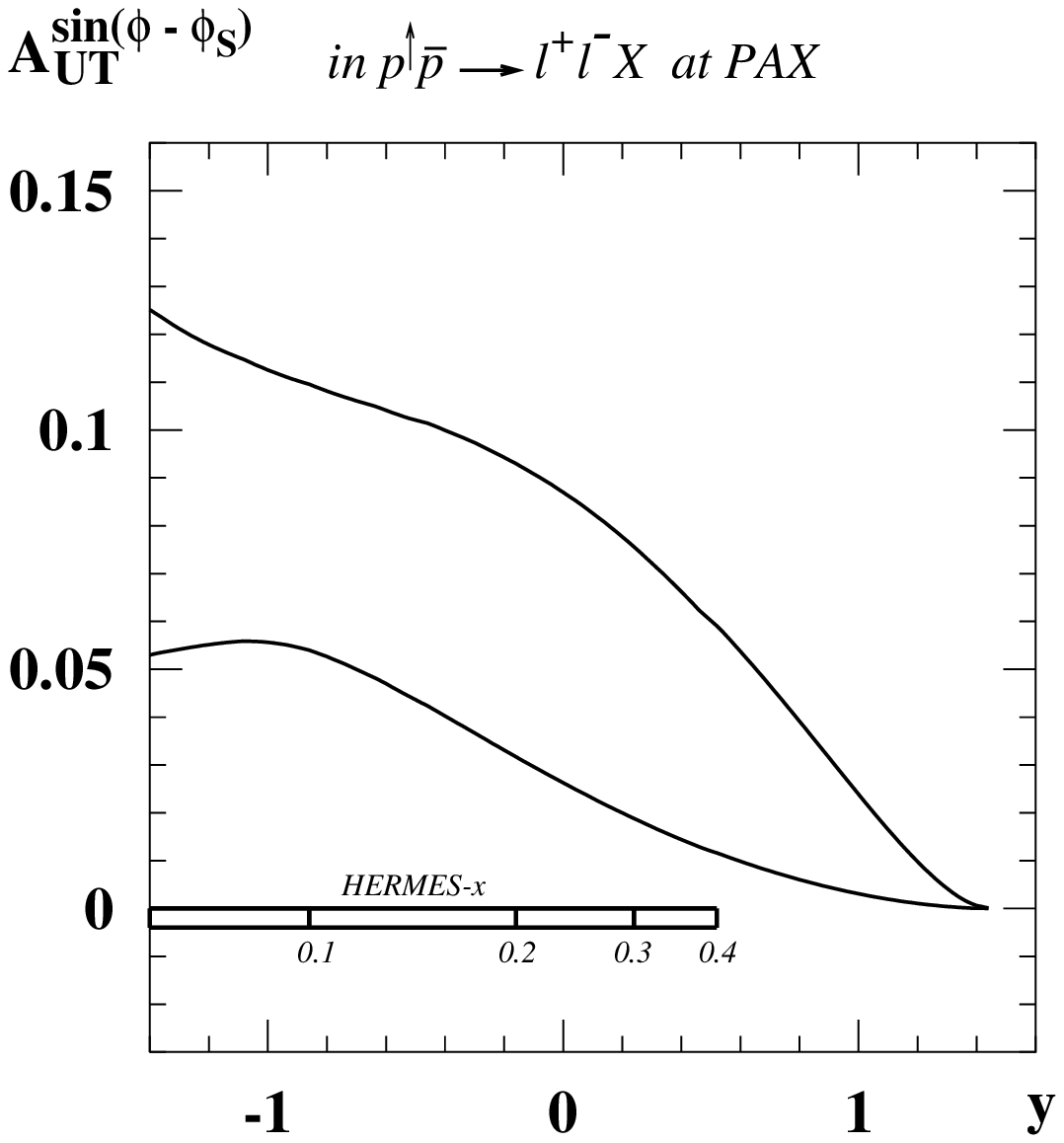}} &
    {\epsfxsize=2.4in\epsfbox{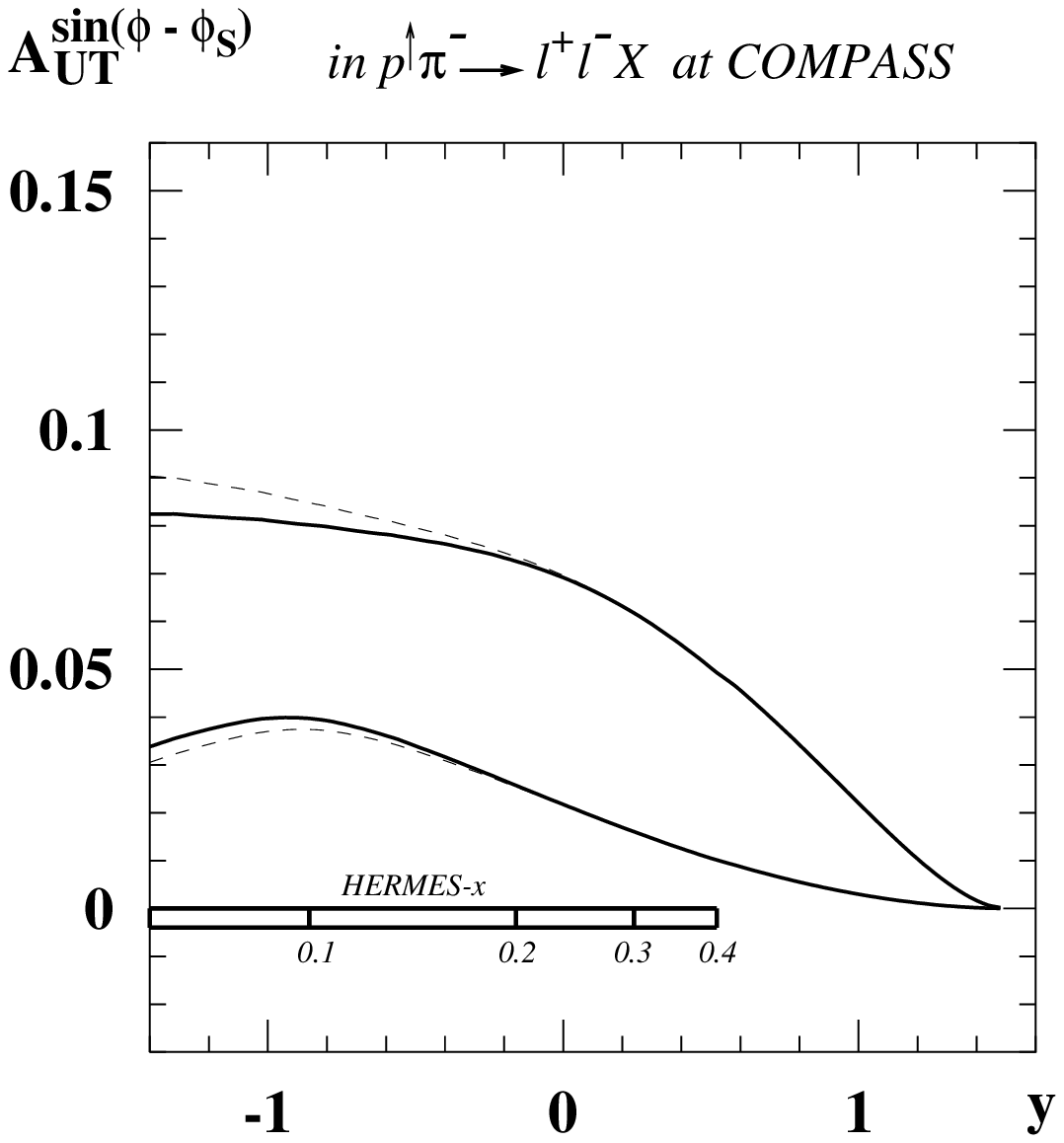}}
\end{tabular}
\caption{\footnotesize\sl
    \label{Fig-DY-at-PAX+COMPASS-new} 
    The azimuthal SSA $A_{UT}^{\sin(\phi-\phi_S)}$ as function of $y$ in
    Drell-Yan lepton pair production in $p^\uparrow \bar{p}\to l^+l^- X$
    for the kinematics of the PAX fixed target experiment
    with $s=45\,{\rm GeV}^2$ and $Q^2=2.5\,{\rm GeV}^2$, and
    in Drell-Yan lepton pair production in $p^\uparrow\pi^-\to l^+l^- X$
    for the kinematics of the COMPASS experiment
    with $s=400\,{\rm GeV}^2$ and $Q^2=20\,{\rm GeV}^2$, respectively.
    The estimates are based on the fit
    for the Sivers $q$-distribution functions, see Eq.~(\ref{Eq:ansatz+fit}),
    obtained from the HERMES data \cite{Airapetian:2004tw}. 
    The inner error band (solid lines) shows the 1-$\sigma$ uncertainty
    of the fit. The outer error band (dashed lines) 
    arises from assuming that the Sivers $\bar q$-distribution functions 
    are proportional to the unpolarized antiquarks see 
    Eqs.~(\ref{Eq:model-Sivers-qbar},~\ref{Eq:model-Sivers-qbar2}).
    For the PAX experiment the uncertainty due to Sivers antiquarks 
    is not visible on this scale.
    The $x$-region explored in the HERMES kinematics is shown. }
\end{figure*}
%

The above numbers are strongly model dependent. The quantitative
results change already by choosing different parameters within the Gaussian
model, cf.\  footnote~\ref{Footnote-choose-diiefferent-parameter}.
However, as an important qualitative conclusion of this exercise,
we point out that in principle the introduction of an appropriately
optimized low-$q_T$ cut may be a helpful device to improve the experimental
signal for the Sivers SSA in DY. 

Since in this context the crucial feature is the presence of a kinematical
zero, the same procedure may turn out helpful in studies of other SSA
or other phenomena related to parton transverse momenta in DY, SIDIS
and other processes.

\section{Sivers SSA in DY at PAX and COMPASS}
\label{Sec-5:Sivers-in-DY-at-PAX-and-COMPASS}

Previously transverse momentum weighted Sivers SSA 
were estimated for the PAX and the COMPASS (hadron mode)
experiments \cite{Efremov:2004tp} on the basis of a 
Sivers function fit to the {\sl preliminary} HERMES data
on the $P_{h\perp}$-weighted Sivers SSA \cite{HERMES-new}.
Here, we solidify the estimates and conclusions of 
Ref.~\cite{Efremov:2004tp} on the basis of the Sivers function fit 
(\ref{Eq:ansatz+fit}) to the {\sl final} HERMES data
\cite{Airapetian:2004tw}.

The fixed target mode at PAX would make available $s=45\,{\rm
  GeV}^2$, which would allow to access lepton pairs in 
$p^\uparrow\bar{p}\to l^+l^-X$ in the 
region $Q^2=2.5\,{\rm GeV}^2$ below the $J/\psi$-resonance and well 
above the region of dileptons from $\phi(1020)$ decays. This $Q^2$
corresponds to the average scale of the HERMES experiment 
\cite{Airapetian:2004tw}, thus at PAX it
is unnecessary to consider $p_T$-broadening effects as we have 
modeled in Eq.~(\ref{Eq:broadening}).

On the basis of the fit (\ref{Eq:ansatz+fit}) we 
obtain for the PAX experiment the estimates plotted 
in Fig.~\ref{Fig-DY-at-PAX+COMPASS-new}.
Nearly the entire range of $y$ for this kinematics
is constrained by results from HERMES data. 
The SSA is sizeable, (5-10)$\%$ in the central region,
and could be measured, see \cite{PAX}. 
The uncertainty due to the unknown Sivers antiquark distribution 
functions is completely negligible, as expected \cite{Efremov:2004tp}.

In the COMPASS experiment, using the possibility of a hadron beam,
one could study SSA in $p^\uparrow\pi^-\to l^+l^-X$ at $s=400\,{\rm GeV}^2$
and, e.g., $Q^2=20\,{\rm GeV}^2$ well above the resonance region of 
$J/\psi$ and other charmonia. In this kinematics $p_T$-broadening 
effects can be expected to be comparable to RHIC, and we use
estimates analogous to (\ref{Eq:broadening}).

Our estimate for the COMPASS experiment is shown in 
Fig.~\ref{Fig-DY-at-PAX+COMPASS-new}. 
Again we observe that for this kinematics one probes the 
Sivers function in the $x$-region, where it is well constrained
by the HERMES data \cite{Airapetian:2004tw}. Also at COMPASS, 
the Sivers SSA is sizeable, about $(4-8)\%$ in the central region,
i.e.\ sufficiently large to allow a test of the prediction (\ref{Eq:01}).  
In order to see this, let us make the following crude estimate.  
The differential cross section for $\pi^-p\to\mu^+\mu^-X$ 
is about $\di\sigma/\di Q \approx 50\,{\rm pb}/{\rm GeV}$
in the region $y>0$ in a kinematics comparable to
that considered here \cite{Anderson:1979tt}.  Assuming an
integrated luminosity of $10^{39}{\rm cm}^{-2}$, which is realistic
\cite{COMPASS-proposal}, one could measure with an ideal detector
and a target polarization of $P=100\%$ the SSA with an accuracy of
$\,\delta A = 1/(P\sqrt{N})\sim 0.5\%$ in the forward region $y>0$
in a bin of size $\Delta Q \sim 1\,{\rm GeV}$.  Of course, this
accuracy cannot be achieved with a realistic detector and
polarization --- but this is not necessary for a first test of the
change of sign in Eq.~(\ref{Eq:01}).

The impact of Sivers antiquark distributions is weak 
and negligible as noted in Ref.~\cite{Efremov:2004tp}. 
In Fig.~\ref{Fig-DY-at-PAX+COMPASS-new} we demonstrate this 
for the model II in Eq.~(\ref{Eq:model-Sivers-qbar2}).
For the estimates in Fig.~\ref{Fig-DY-at-PAX+COMPASS-new}
we used the parameterizations \cite{Gluck:1998xa,Gluck:1999xe}
(using \cite{Martin:2002dr,Sutton:1991ay} confirms the results).

Thus, though at the present stage one cannot provide dedicated estimates of $\delta A$ 
for these experiments as we did for RHIC in Sec.~\ref{Sec-3:Sivers-in-DY-at-RHIC}, 
we conclude that also PAX and COMPASS could test the prediction (\ref{Eq:01}) for 
quarks. However, both experiments are insensitive to Sivers antiquark distributions 
\cite{Efremov:2004tp}, which underlines the unique feature of RHIC 
with respect to this point.

We stress that quantitative analyses and extractions of the Sivers function 
from COMPASS, PAX and RHIC data will require a good theoretical control on 
NLO-QCD-corrections, soft factors and evolution effects.
For discussions of some of these issues in the context of (collinear) 
double spin asymmetries in particular for the PAX kinematics see 
\cite{Ratcliffe:2004we,Shimizu:2005fp}.

\section{Conclusions}
\label{Sec-6:Summary}

We conclude that PHENIX and STAR can confirm the change of sign of the
Sivers $q$-distribution function in SIDIS and the Drell-Yan process in
Eq.~(\ref{Eq:01}) (as can PAX with an antiproton and COMPASS with 
a pion beam). Both PHENIX and STAR can provide first information on the 
Sivers $\bar q$-distribution functions which are not constrained by the 
current SIDIS data from HERMES or COMPASS, unlike PAX and COMPASS.

There are more recent (preliminary!) HERMES data
\cite{Diefenthaler:2005gx}. Our fit is compatible with the new data,
but the tendency is to increase the Sivers function. This tendency
would result in even more optimistic estimates for DY at RHIC (and COMPASS
and PAX).

We confirm Refs.\ \cite{Anselmino:2005ea,Vogelsang:2005cs} with
respect to the capability of RHIC to access the Sivers
$q$-distribution.  In addition we point out the possibility to learn
about Sivers $\bar q$-distribution from RHIC.

\begin{acknowledgments}
The work is partially supported by BMBF and DFG of Germany, the
COSY-Juelich project, the Transregio Bonn-Bochum-Giessen, and is
part of the European Integrated Infrastructure Initiative Hadron
Physics project under contract number RII3-CT-2004-506078. 
A.~E.\ is supported by Grant No. RFBR 06-02-16215 and 
by a special program of the RF MSE Grant No. RNP.2.2.2.2.6546.
JCC is supported in part by the US DOE.
\end{acknowledgments}


\end{document}